\documentclass[aps,twocolumn,prb,superscriptaddress,amsmath,tightenlines]{revtex4}

\usepackage{graphicx}
\begin{document}

\title{Simulations of hybrid charge-sensing single-electron-transistors and CMOS circuits
}

\author{Tetsufumi Tanamoto}
\affiliation{Department of Information and Electronic Engineering, Teikyo University,
Toyosatodai, Utsunomiya  320-8551, Japan} 
\email{tanamoto@ics.teikyo-u.ac.jp}

\author{Keiji Ono}
\affiliation{Advanced device laboratory, RIKEN, Wako-shi, Saitama 351-0198, Japan}

\begin{abstract}
Single-electron transistors (SETs) have been extensively used as charge sensors
in many areas such as quantum computations. 
In general, the signals of SETs are smaller than those of complementary metal-oxide semiconductor (CMOS) devices, 
and many amplifying circuits are required to enlarge these signals. Instead of amplifying a single small output, 
we theoretically consider the amplification of pairs of SETs, such that one of the SETs is used as a reference. 
We simulate the two-stage amplification process of SETs and CMOS devices using a conventional 
SPICE (Simulation Program with Integrated Circuit Emphasis) circuit simulator.
Implementing the pairs of SETs into CMOS circuits makes the integration of SETs more feasible 
because of direct signal transfer from the SET to the CMOS circuits.

\end{abstract}

\maketitle

Single-electron transistors (SETs) have been intensively investigated owing to their advantages of low-power operation, 
which is desirable for application to logic and memory elements~\cite{Grabert, Likharev}. 
Since Tucker's proposal~\cite{Tucker}, many approaches have been developed to replace the elements 
in complementary metal-oxide semiconductor (CMOS) circuits~\cite{Chen,Tiwari,Yano,Mahapatra,Inokawa,Uchida}.
SETs were also investigated after they were directly embedded in CMOS circuits~\cite{Yano,Uchida2,Inokawa}. 
Currently, conventional Si transistors are much smaller than SETs. 
SETs have attracted attention as charge sensors~\cite{Field,Berman,Ionescu}, 
which have been used for the readouts of silicon qubits~\cite{Morello,Gonzalez-Zalba,Shaji}, 
or as current standards~\cite{Fujiwara,Nishiguchi,Fujiwara2}.

The SET consists of a small metallic island or a degenerated semiconductor island surrounded by a source 
and drain via tunneling junctions with low capacitances. The SET yields periodic outputs referred to as Coulomb oscillations,
which vary as a function of the gate voltage, and the oscillation period corresponds 
to the change in the number of electrons in the island. 
The maximum signal change in an SET is the difference between the peak and the trough of the Coulomb oscillation. 
The SET current is sensitive to electric potential variations on the SET island~\cite{Field}. 
The standard readout process of a spin qubit (spin in localized states, such as
a single impurity or quantum dot (QD)) is a spin-to-charge conversion, 
and the change in charge of the localized state is detected by the SET current~\cite{Schoelkopf,Lu,Morello,Gonzalez-Zalba}.

There are many methods for the detection of charges, such as the use of 
radiofrequency SETs (rf-SETs)~\cite{Schoelkopf,Lu,Gonzalez-Zalba}
and the direct measurement of the SET by a single GaAs amplifier~\cite{Petersson,Visscher}. 
Inokawa {\it et al} showed that a Coulomb oscillation is effectively outputted by directly connecting the SET 
to an MOS field-effect transistor. They experimentally demonstrated multiple-valued logic by SETs operating at 27 K. 
The effectiveness of the series coupling of the SET with MOS transistors was also experimentally investigated 
by Uchida {\it et al.}~\cite{Uchida2}. 
Scalable SET sensing systems require scalable circuits. 
However, previous SET sensors~\cite{Morello,Gonzalez-Zalba,Shaji,Fujiwara,Nishiguchi,Fujiwara2} 
did not explicitly consider the array of SETs as part of CMOS circuits.

\begin{figure}
\centering
\includegraphics[width=8.5cm]{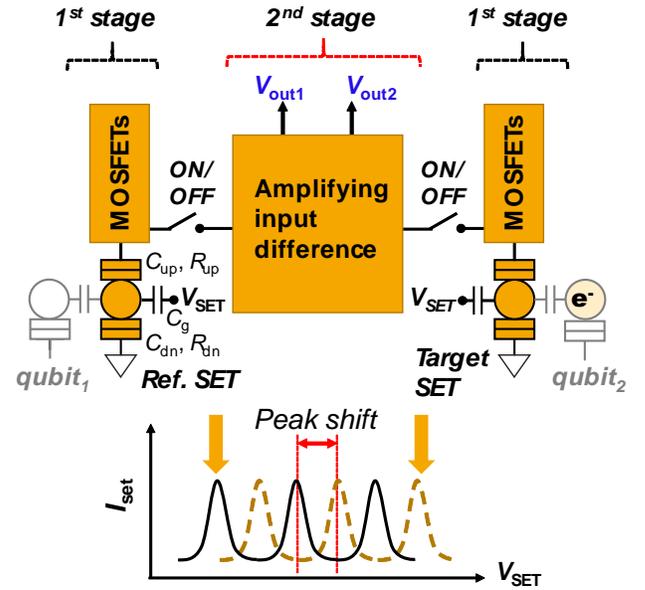}
\caption{
Concept of our two-stage amplification circuit of the charge-sensing single-electron-transistor (SET).
Instead of amplifying one SET sensor, 
we consider amplifying pairs of SETs that have different states with each other. 
Depending on the existence of the extra electron in the quantum dot (QD) outside the SETs, 
the electric potential of the island of the SETs changes.
Throughout this paper, we model the effects of the electric potential of the QDs and $V_{\rm SET}$ 
by the gate voltages $V_{\rm G1}$ and $V_{\rm G2}$ of the SETs. The switches between the 1st and 2nd stages 
are typically realized by wordline nMOS transistors, using which we can select pairs of SETs among the SET array. }
\label{fig1}
\end{figure}

In this study, we theoretically consider a scalable detection circuit based on the implementation of many SETs 
in CMOS circuits. Our basic concept of the amplification of the SET signals consists of two stages, as shown in Fig.~1. 
The first amplification stage of the process is conducted by direct connection of the SET to pMOS transistors. 
In the second stage, we introduce a reference and a target SET 
and amplify the difference between two SETs using standard amplifier circuits, 
such as the differential amplifier (DA) circuits, and the static random-access memory (SRAM) cells.

For smooth connection to the digital circuits, the output of the sensing circuit should be a digital signal (either a "0" or a "1").
For this purpose, it is better to compare the relative output of the target SET with that of the reference SET. 
In our application of the SRAM cell, the relative voltage difference between the target SET and the reference SET is found 
to be quickly latched to "0" or "1.” This is in contrast to the measurements in Refs.~\cite{Morello,Gonzalez-Zalba,Shaji}, 
wherein the results were obtained after the analysis of a series of time-dependent SET currents. 
Our CMOS circuits are assumed to be close to the SETs. 
In addition, 
the target and the reference SETs are chosen by switching on the wordline transistors between the pairs of SETs and amplifiers. 
Then, the circuit using the pairs of SETs becomes more compact than the amplifying circuit of a single SET. 
Accordingly, our proposal is suitable for an array of sensing SETs.

In this study, we implement the current characteristics of the orthodox theory of the SET~\cite{Amman} 
into the SPICE (Simulation Program with Integrated Circuit Emphasis) circuit simulator 
based on the BSIM4 (Berkeley Short-channel transistor Model, level=54)
by using the standard modeling language of the Verilog-A.
We considered two types of SETs one of which is operated at low (4.2 K), and the other is operated at high temperatures (-30$^\circ$C). 
At low temperatures, such as 4.2 K, the threshold voltage of the MOS transistors becomes higher because of the incomplete ionizations.
Many studies regarding cryo-CMOS~\cite{Green,Gaensslen,Beckers,Dijk} have been conducted to determine the model parameters at low temperatures. 
However, the general compact model is not available for such low temperature. 
In addition, the basic CMOS operations are basically the same as that at room temperature (RT).
Thus, we applied the CMOS parameters that are available in the conventional SPICE models to the CMOS parts 
for both types of the SETs. 
In the following, circuit calculations are mainly conducted at $T= -30^\circ$C, 
which is in the range of the conventional models.

We consider CMOSs with gate lengths of $L=$90 nm and $L=$65 nm, whose drain voltages $V_{\rm D}$ are 1.2 V and 1.0 V, respectively. 
We also consider the effects of small variations in the SETs and CMOS transistors. 
The purpose of the comparison of pairs of SETs is to detect the changes in the Coulomb oscillations of the target SET.
Thus, given that the voltage difference between the peak and trough currents of the Coulomb oscillations 
can be distinguished, we will be able to 
detect the changes of the target SET even if there are variations in the devices.

\begin{table}
\caption{
The four SETs we use here.
}
\label{table:SET}
\begin{center}
\begin{tabular}{l|l|l|l|l|l|l|l}\hline
&\multicolumn{3}{|c|}{Capacitance(aF)} & \multicolumn{2}{|c|}{Resistance($\Omega$)} & Temp(K) & $E_c$(meV)\\ \hline
 SET & $C_{\rm up}$ & $C_{\rm dn}$  & $C_g$  & $R_{\rm up}$  & $R_{\rm dn}$  & $T$ & \\ \hline
LT-SET$r$ 
& 1  & 10 & 2 & 100k & 1M  & 4.2 & 6.16\\ \hline
LT-SET$t$  
& 1.1 & 0.9 & 2.2 & 110k & 0.9M  &4.2 & 6.51\\ \hline
RT-SET$r$  
& 0.1 & 0.5   & 0.5 & 100k & 500k & 243 & 72.8 \\ \hline
RT-SET$t$  
& 0.11 & 0.45   & 0.45 & 110k & 450k & 243 & 79.3 \\ \hline
\end{tabular}
\end{center}
\end{table}

\begin{figure}
\centering
\includegraphics[width=8.5cm]{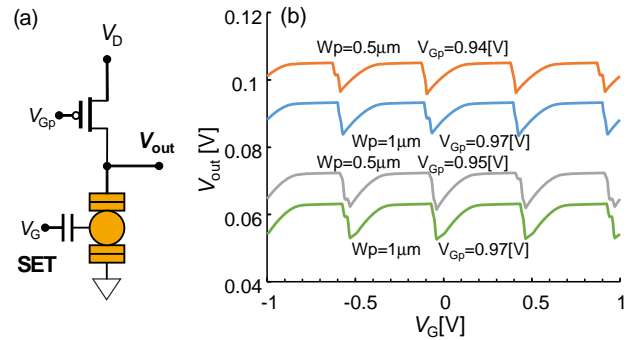}
\caption{
First-stage amplification circuits, 
where the LT-SET$t$ (see table I) and pMOS are directly connected. The output voltage $V_{\rm out}$ is amplified depending on the gate voltage $V_{\rm Gp}$ of the pMOS. 
Depending on the width of the pMOS ($W_p =$ 0.5 $\mu$m and $W_p =$ 1 $\mu$m), 
the optimal points change. The Coulomb oscillation is amplified up to approximately 10 mV ($V_{\rm D} = 1.2$ V). 
Hereafter we use $W_p =$ 0.5 $\mu$m pMOS at the 1st stage.}
\label{figpmos1}
\end{figure}

Herein, we consider four SETs, as listed in Table I. The LT-SET$t$ (target) and LT-SET$r$ (reference) are operated at 4.2 K, 
and RT-SET$t$ and RT-SET$r$ can operate at -30$^\circ$C. 
The parameters of the LT-SET$t$ (RT-SET$t$) exhibit 10\% variations compared with those of the LT-SET$t$ (RT-SET$r$). 
The calculated current–voltage characteristics ($I_{\rm D}$-$V_{\rm D}$) are listed in the appendix. 
The charging energy is given by $E_c \equiv e^2 / [2(C_{\rm up} + C_{\rm dn} + C_{\rm g})]$. 
The magnitude between the peak and trough currents is on the order of pA, 
and its voltage change estimated by pA$\times$ 25.9 k$\Omega =$ 25.9 $\mu$V is very small 
(25.9 k$\Omega$ is a quantum resistance~\cite{Grabert}). 
On the other hand, the variations in the threshold voltage $V_{\rm th}$ of conventional 
CMOSs are generally on the order of millivolts. 
Therefore, we need to amplify the SET outputs before connecting the SETs to conventional CMOS circuits.  

\begin{figure}
\centering
\includegraphics[width=7.5cm]{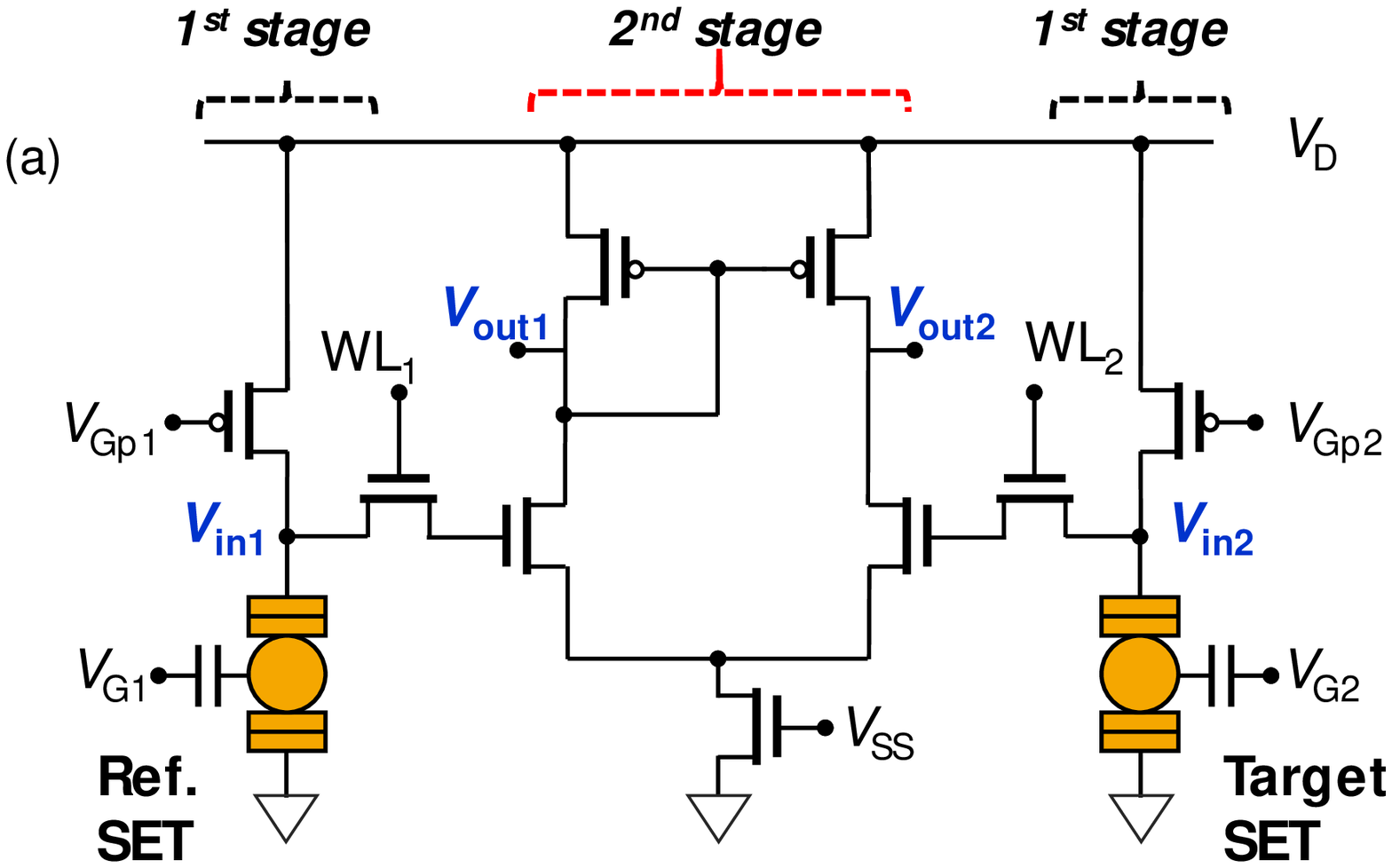}
\includegraphics[width=6.8cm]{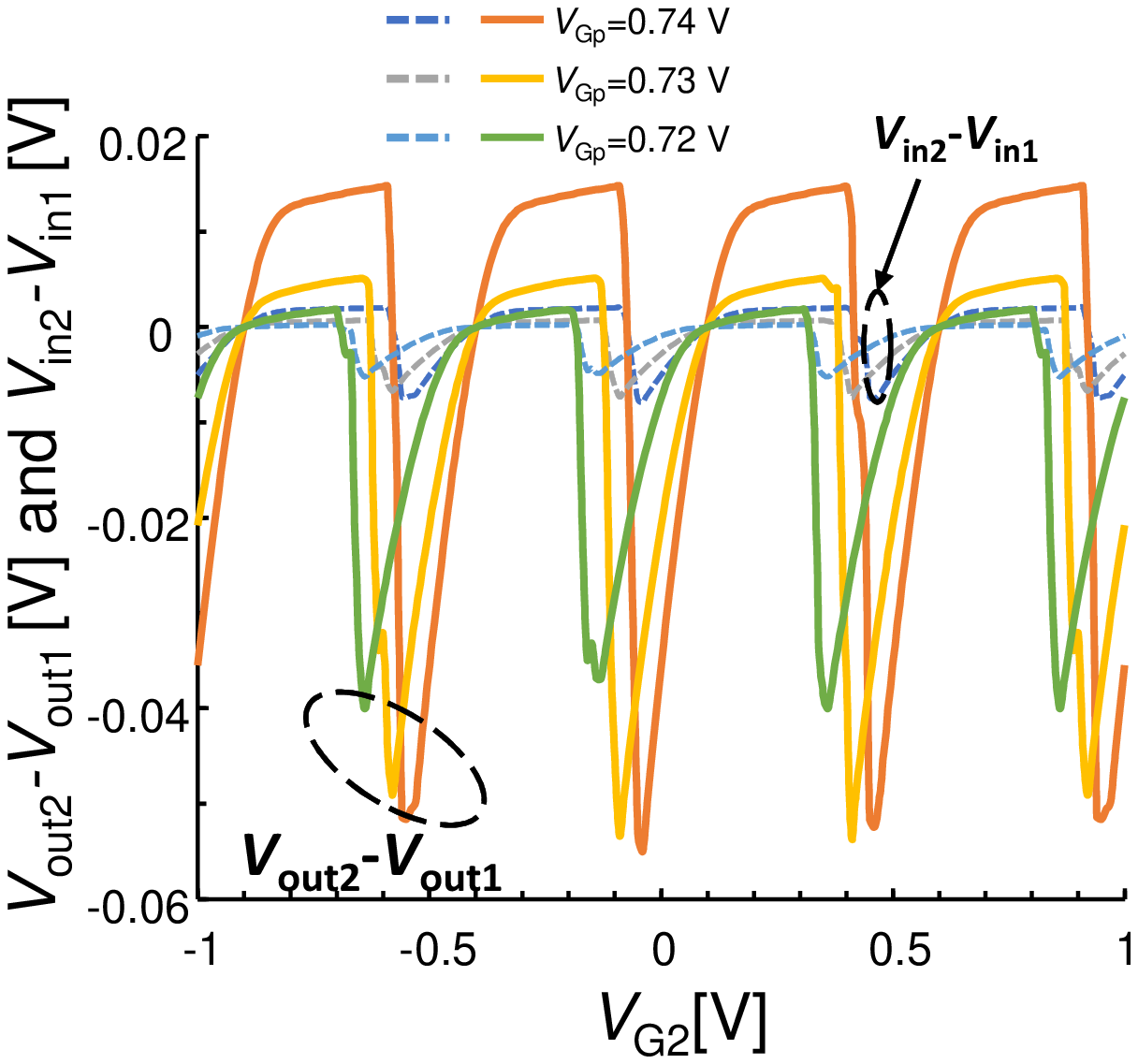}
\includegraphics[width=7.5cm]{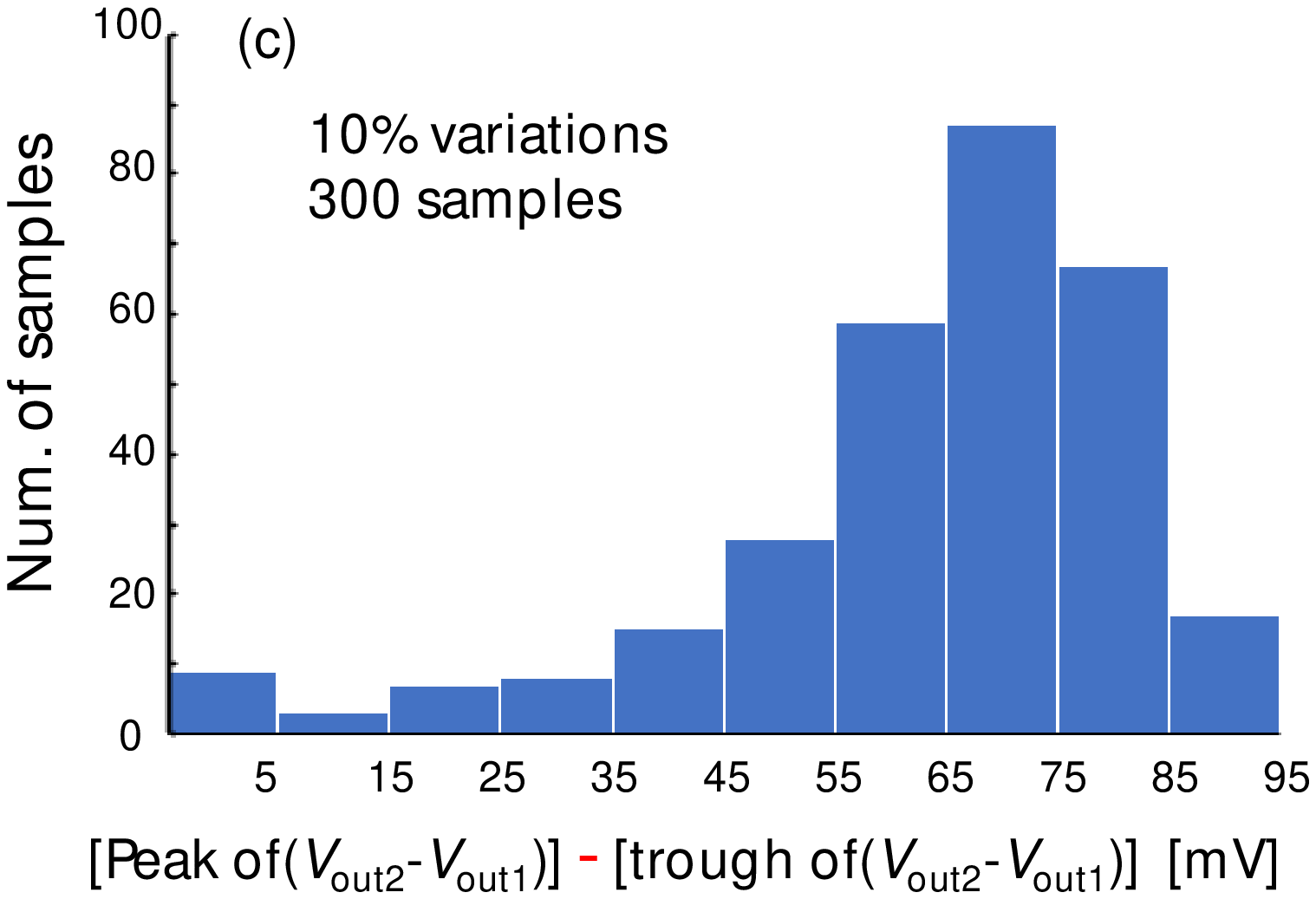}
\caption{
(a) The differential amplifier (DA) is applied at the 2nd stage amplification of Fig.1.
(b) Numerical results of $V_{\rm out1}$-$V_{\rm out2}$ 
(enhanced Coulomb oscillations) for $L=$90 nm at $V_{\rm G1}=0$. 
$V_{\rm Gp1}=V_{\rm Gp2}=V_{\rm Gp}$.
The pMOS and nMOS widths of the DA are given by
$W_{\rm pa} = 1 \mu$m, and $W_{\rm na} =$10 $\mu$m, respectively.
The widths of the wordline nMOS and 1st stage pMOS are given by $W_{\rm n} = 5$ $\mu$m and $W_{\rm p} =$ 0.5 $\mu$m, respectively.
$V_{\rm D}=1.2$ V.
We can see that the amplitude of $V_{\rm out2}-V_{\rm out1}$ is approximately 40 mV, which is 
larger than the amplitude of the input $V_{\rm in2}-V_{\rm in1}$.
(c) Histogram of the Monte Carlo simulation of the output difference 
when the $V_{\rm th}$ of all CMOS transistors varies by 10\% over 300 simulations.}
\label{figdiff}
\end{figure}

We start from the first amplification stage in which the SET is directly connected to the CMOS transistors, as shown in Fig.~\ref{figpmos1}(a). By a series connection, the low SET current value increases with CMOS~\cite{Inokawa}. 
Figure~\ref{figpmos1}(b) shows that the amplification of the SET signal becomes prominent at $V_{\rm Gp}\gtrsim 0.9$ V. 
The distortions of the waveforms compared with the original Coulomb oscillations originate from the nonlinearity of the $I_{\rm D}$-$V_{\rm D}$ characteristics of the pMOS transistors. 
We analyze the amplification mechanism based on the standard long-channel model~\cite{Taur}. 
The $I_{\rm D}$-$V_{\rm D}$ of CMOS transistors depends on the triode region $|V_{\rm DS}| < |V_{\rm GS} - V_{\rm th}|$ and the saturation region $|V_{\rm DS}| > |V_{\rm GS} - V_{\rm th}|$. 
At present, amplification is observed in $|V_{\rm DS}| = |V_{\rm out}-V_{\rm D}|\approx $1.05 V$ > |V_{\rm GS}|=|V_{\rm Gp} - V_{\rm D}|= 0.15$ V (saturation region). We did not observe a sufficiently large amplification in the triode region. 
The SET current $I_{\rm SET}$ under large source-drain voltage is approximately described by $I_{\rm SET} \propto V_{\rm SET}$ (see appendix). 
More explicitly we assume that $I_{\rm SET} \approx V_{\rm SET}/R_{\rm D}$.
Thus, we can write the $I_{\rm D}$-$V_{\rm D}$ characteristics in the saturation region; further, the SET given by 
\begin{eqnarray}
I_{\rm D} &=& \frac{1}{2}\beta_p (V_{\rm Gp}-V_{\rm D}-V_{\rm thp})^2 (1+\lambda (V_{\rm D}-V_{\rm out})), \label{eq1-1} \\
V_{\rm out}&\approx&R_{\rm D} I_{\rm D}. \label{eq1-2}
\end{eqnarray}
The solution of these equations is given by
\begin{equation}
V_{\rm out}=\frac{1+\lambda V_{\rm D}}{1+\lambda I_{D0} R_{\rm D}} I_{D0} R_{\rm D},  
\label{eq1out}
\end{equation}
where $I_{D0}\equiv\frac{1}{2}\beta_p (V_{\rm Gp}-V_{\rm D}-V_{\rm thp})^2$.
Because $I_{D0} R_{\rm D} <V_{\rm D}$, the SET output is enhanced by 
$\frac{1+\lambda V_{\rm D}}{1+\lambda I_{D0} R_{\rm D}} >1$.
Note that $\lambda$ is conspicuous in both the $L=$90 nm and $L=$65 nm pMOS transistors (see appendix), 
and we can see that the output voltage oscillates around $\sim$10 mV in Fig.~\ref{figpmos1}(b).


\begin{figure}[h]
\centering
\includegraphics[width=8.0cm]{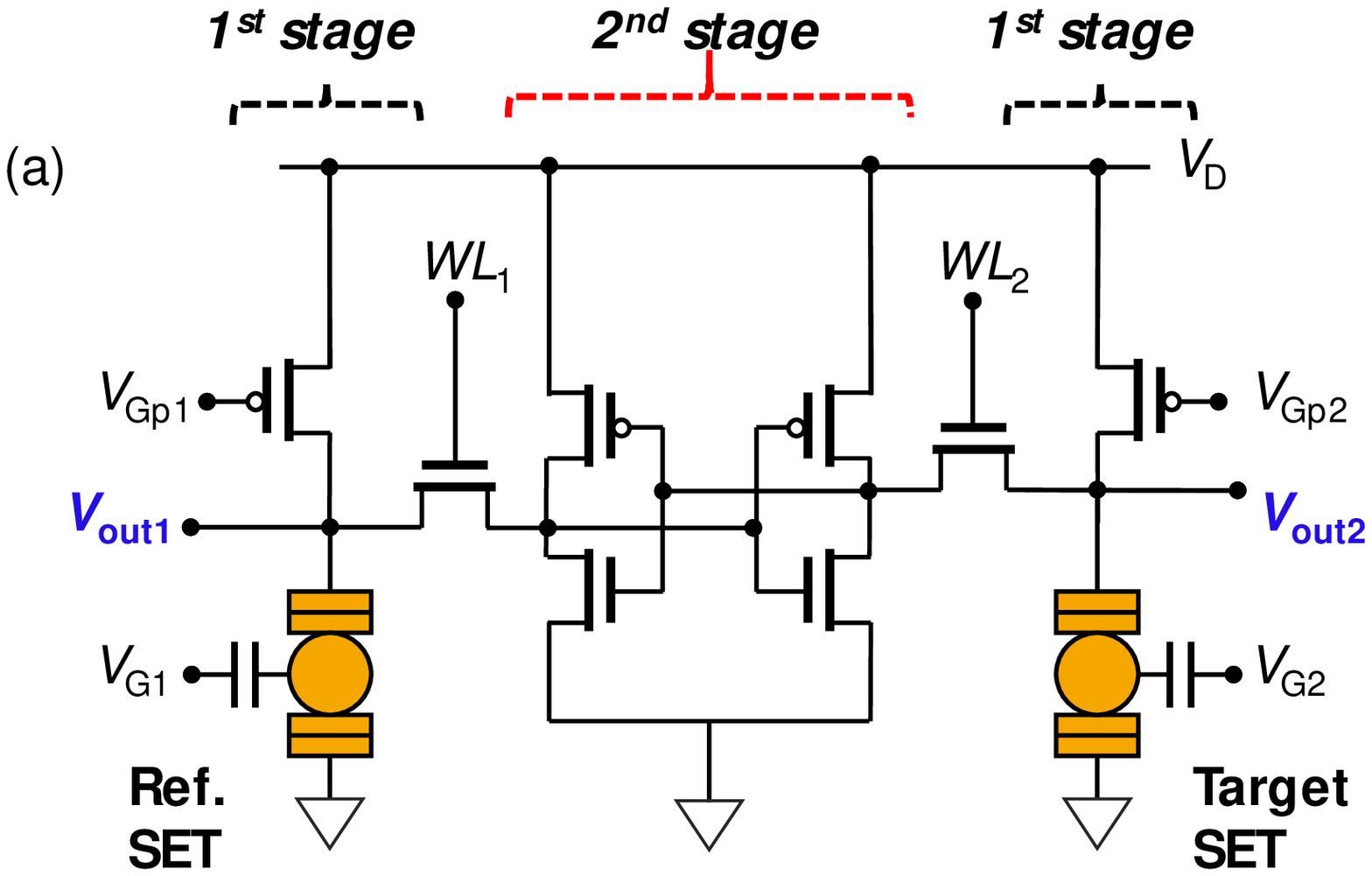}
\includegraphics[width=8.8cm]{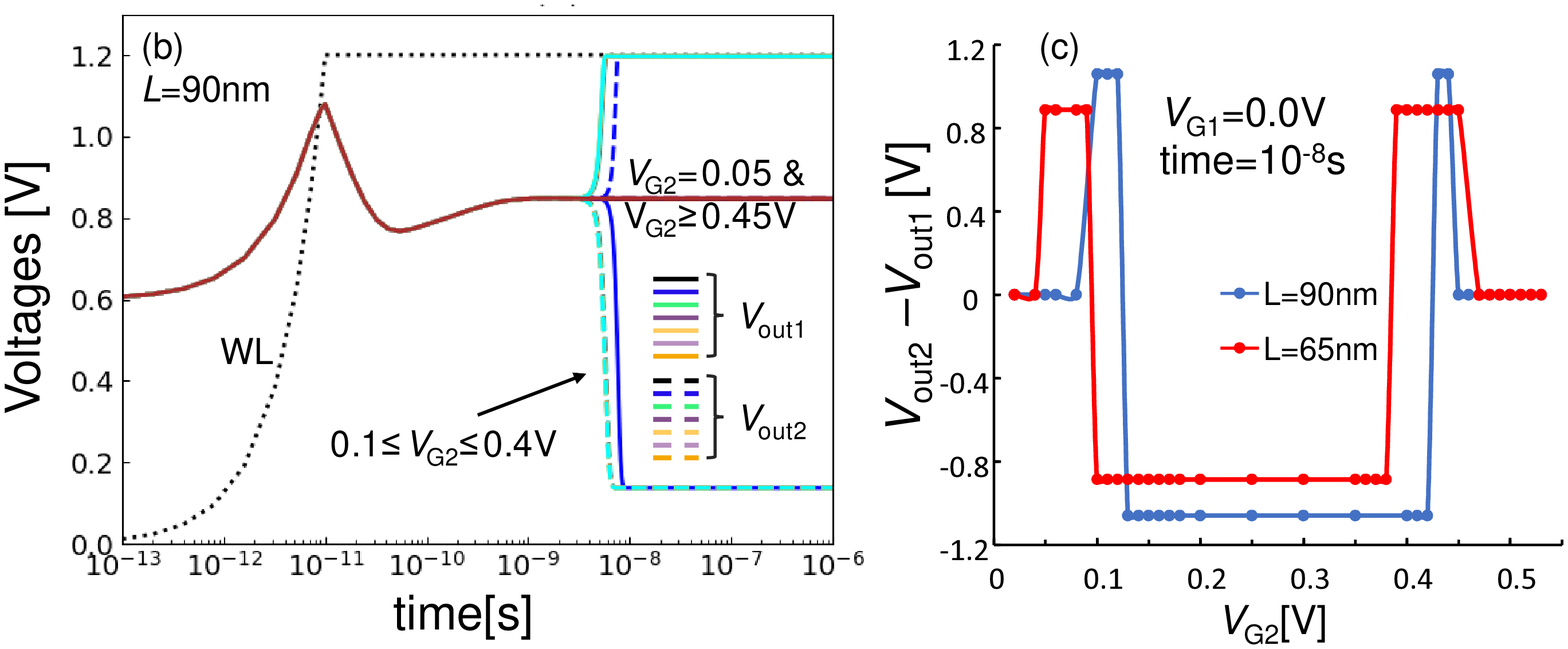}
\caption{(a) Six transistor  static random-access memory (SRAM) cell applied in the 2nd-stage amplification of Fig.~1.
(b) Time dependent voltage behaviors of the SRAM setup of $L$=90 nm for widths $W_n=0.5$ $\mu$m and  $W_p =1.2 W_n$.
The width of the $WL$ transistor is 0.4 $\mu$m. 
$V_{\rm Gp}=0.55$ V and $V_{\rm G1}=0.0$ V using the LT-SET$t$s.
The change in $V_{\rm out1}$ and $V_{\rm out2}$ is in the range of 
$V_{\rm G2}=\{$0, 0.05, 0.1, 0.15, 0.2, 0.25 0.3, 0.35, 0.4, 0.45, 0.5 $\}$ V.
(c) Replotting of (b) as a function of $V_{\rm G2}$ of (b) and the result for $L=$65 nm.}
\label{figSRAM}
\end{figure}
\begin{figure}[h]
\centering
\includegraphics[width=8.8cm]{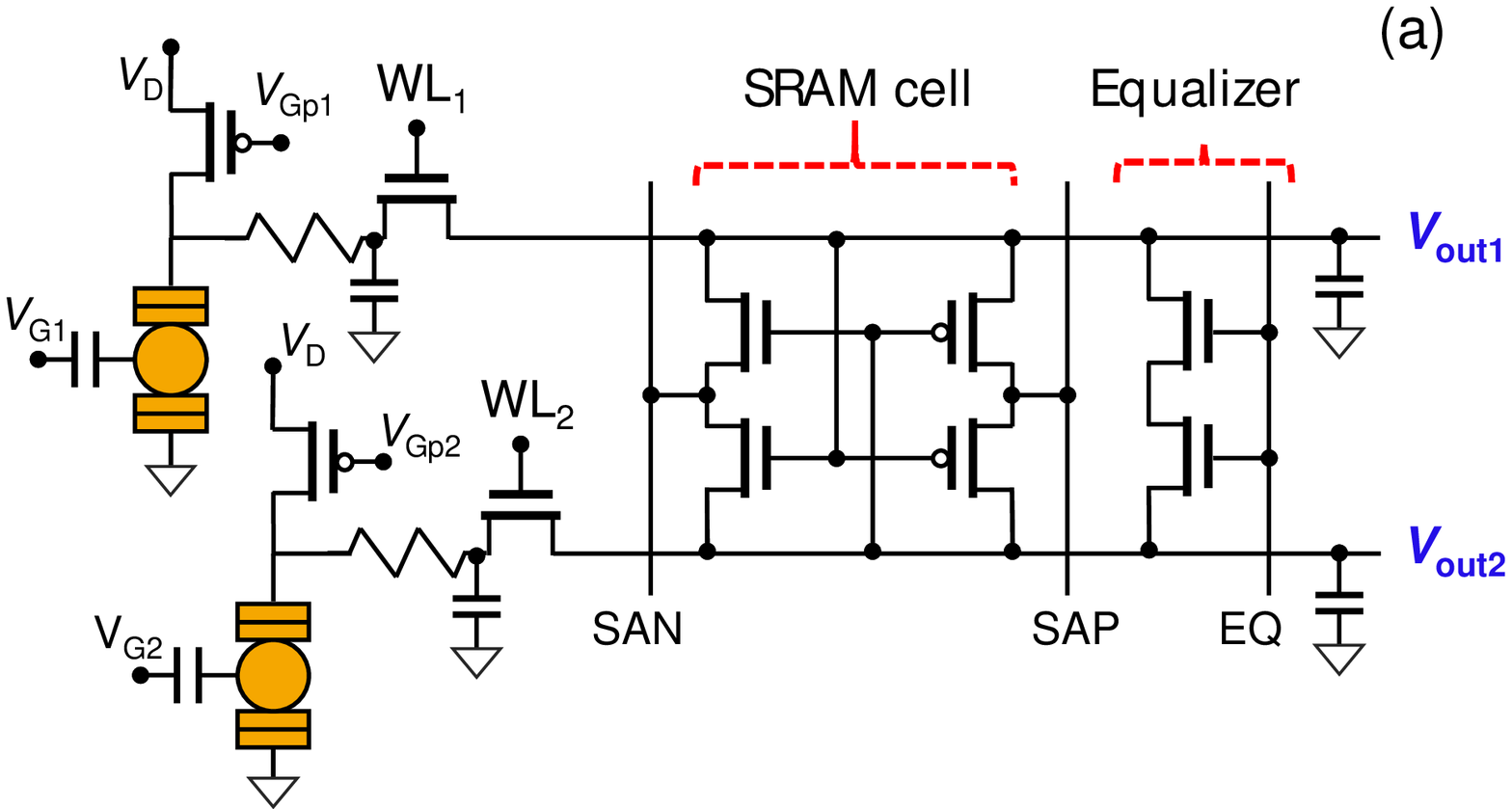}
\includegraphics[width=8.8cm]{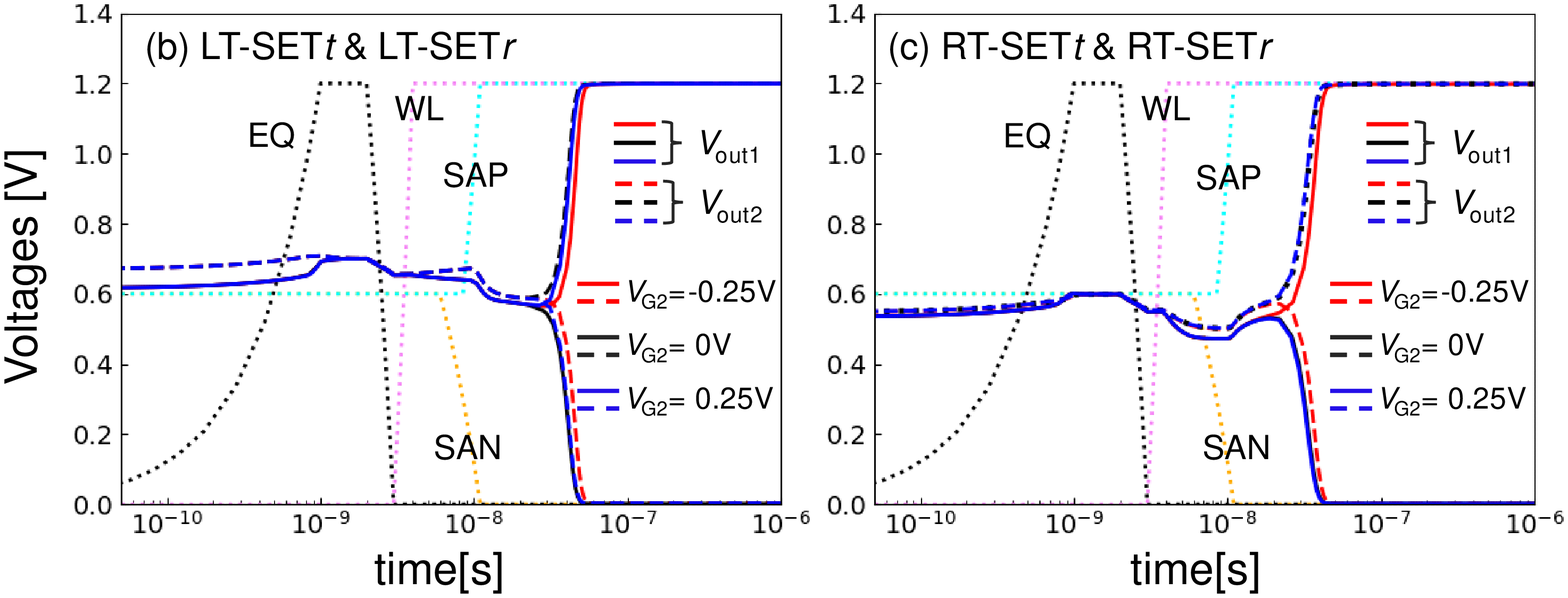}
\caption{
(a) Dynamic RAM(DRAM)-like detection circuits are applied at the 2nd stage amplification of Fig.~1.
The stray capacitance of 1 pF is added to the nodes $V_{\rm out1}$, $V_{\rm out2}$, 
inputs of the SETs, and equalizer circuits.
(b)(c) Time-dependent characteristics of the different SET pairs (the LT-SET pair in (b) and 
RT-SET pair in (c)) in Fig.~(a)
with 10\% threshold voltage variations in the MOS transistors.
The pulse sequence is constituted following the standard DRAM sequence of Ref.~\cite{DRAM1,DRAM2}.
$V_{\rm G1}=0$ is fixed and $V_{\rm G2}$ changes.
Depending on whether $V_{\rm G1}>  V_{\rm G2}$ or $V_{\rm G1}< V_{\rm G2}$, 
the outputs $V_{\rm out1}$ and $V_{\rm out2}$ change from $V_{\rm out1}> V_{\rm out2}$
to  $V_{\rm out1}< V_{\rm out2}$.
The widths of the nMOS and pMOS are 0.5 $\mu$m and 0.6 $\mu$m. The width of the equalizer nMOS is 20 $\mu$m.
The width of the $WL$ transistor is 0.6 $\mu$m.
($V_{\rm Gp1}=V_{\rm Gp2}=0.87$ V, $L=$90 nm). 
}
\label{figDRAM}
\end{figure}

The second-stage amplification is conducted using standard amplifier circuits. 
Figure~\ref{figdiff} shows the results obtained from a basic DA circuit.
Note that the gate voltages $V_{\rm G1}$ and $V_{\rm G2}$ represent the shifted electrical potential $V_{\rm SET}$ by the additional sensing QD in Fig.~\ref{fig1}. 
Conventional DAs amplify two inputs with opposite phases. 
In the case of SETs, two different phases of Coulomb oscillations are input. 
We consider that the two SET signals possessing different current peaks mimic the two input signals of the conventional DA with different phases. 
In Figs.~\ref{figdiff}(b), we show the results for the output voltage difference of $V_{\rm out2}-V_{\rm out1}$ for a fixed $V_{\rm G1} = 0$. The voltage difference $V_{\rm out2} - V_{\rm out1}$ increases by approximately 40 mV for $L = 90$ nm (the results for $L = 65 $nm are shown in the appendix), 
and can be detected by conventional CMOS amplifier circuits 
(the following circuits after $V_{\rm out2}$ and $V_{\rm out1}$ are not shown). 
This enhancement of the Coulomb oscillations is the result of the two-stage amplification of SET signals.

The magnitude of the enhancement of the Coulomb oscillation in Fig~\ref{figdiff}(b) changes 
depending on the threshold voltage variations
of the eight MOS transistors. 
In Fig.\ref{figdiff}(c), we provide the distribution of the difference in 
the peak and trough of $V_{\rm out2}-V_{\rm out1}$
obtained through over 300 Monte Carlo simulations of the threshold variations 
with LT-SET$t$ and LT-SET$r$ at $V_{\rm G1}=0$ and $V_{\rm G2}=0.2$ V.
The number of small amplitudes of $|V_{\rm out2}-V_{\rm out1}|$ 
($<$ 5 meV) is nine out of 300 samples.
Small amplitudes of $V_{\rm out2}-V_{\rm out1}$ can be avoided by applying different voltages such as $V_{\rm Gpi}$ and $WL_i$
on each part of the DA.

We now consider the application of the standard SRAM cell containing six MOS transistors~\cite{Seevinck} to detect a pair of SETs, as shown in Fig.~\ref{figSRAM}(a). 
Figure~\ref{figSRAM}(b) shows the simulation results of the time-dependent SRAM cell outputs $V_{\rm out1}$ and $V_{\rm out2}$ of the two LT-SET$t$s. 
We can see that $V_{\rm out1} < V_{\rm out2}$ for $V_{\rm G2} = 0.1$ V, but $V_{\rm out1} > V_{\rm out2}$ for $0.15$ V$\le V_{\rm G2} \le 0.4 $V for $L$= 90 nm devices. 
This implies that the shift in the electric potential of the sensing QD from $V_{\rm G2} = 0.1$ V to $V_{\rm G2} = 0.15$ V changes the electric potential of the target SET island, resulting in a change in the relative magnitude between $V_{\rm out1}$ and $V_{\rm out2}$. 
Figure~\ref{figSRAM}(c) replots $V_{\rm out2}-V_{\rm out1}$ in Fig.~\ref{figSRAM}(b) as a function of $V_{\rm G2}$ with the results for $L = $ 65 nm at time $10^{-8}$ s. 
Thus, we can detect the change in the target SET represented by $V_{\rm G2}$ by measuring the relative outputs of the SRAM cell. 
Herein, the initial voltage at the SRAM cell input is set to $V_{\rm D}/2$, and stray capacitances of 0.2 pF are included at the input nodes of the SRAM cell (Figures not shown). 
As the stray capacitance increases, the time to split increases.

In general, SRAM cells undergo initial threshold voltage $V_{\rm th}$ variations~\cite{PUF}, 
and we have to consider these variations in the MOS transistors as well as the two SETs.
Here, we extend the SRAM cell circuit to a dynamic random-access memory (DRAM)-like structure~\cite{DRAM1,DRAM2} in Fig.~\ref{figDRAM}(a), where it is considered that the equalizer circuit mitigates the voltage difference of the wordlines between the two SETs.
Figures~\ref{figDRAM}(b)(c) show two types of readouts for the two types of SETs. 
The equalizer voltage is switched on at 1 ns and stopped after 2 ns. 
The SAN and SAP voltages are switched on after the equalizer at $t = 6$ ns and $t = 8.5$ ns. 
The wordlines are switched on at 4 ns. 
We can see that the change in the gate voltage $V_{\rm G2}$, which corresponds to the existence of the charge sensor QD, causes the outputs $V_{\rm out1}$ and $V_{\rm out2}$ to change from $V_{\rm out1} > V_{\rm out2}$ to $V_{\rm out1} < V_{\rm out2}$.

In a realistic situation, it is possible that the temperature changes. 
Therefore, we also calculated the temperature dependence of the amplifier response. 
It is desirable that the relative magnitude of $V_{\rm out1}$ to $V_{\rm out2}$ does not change 
even when the operating temperature changes. 
The temperature dependencies of the characteristics of the DA (Fig.~\ref{figdiff}) 
and DRAM (Fig.~\ref{figDRAM}) are robust against temperature changes 
as long as the temperature change is sufficiently small on the order of several tens of degrees 
(see appendix).

Considering that there are variations in SETs and MOS transistors, 
we may have to check and record the basic characteristics of each device at the first calibration stage of the chip. 
The $I$–$V$ characteristics of each SET should be clarified before using the charge-sensing SET. 
This information is stored in the extra SRAM or other memories, and becomes the overhead of the system. 
An effective method for determining the optimal biases automatically is a future issue.
It is possible that the applied voltages destroy the SET charge states. 
However, herein, we neglected the effects of the backaction from the CMOS circuits. 
Hence, the assessment of the backaction remains a future issue.

In conclusion, we proposed two-stage amplification circuits for SETs. 
Based on serial connections with MOS transistors and a comparison with the reference SET, 
we numerically show that the readout of the charge-sensing SET is greatly enhanced. 
We also considered the effects of variations in the MOS transistors and SETs, 
and show that as long as the variations are small, the two SETs can be compared effectively.

\subsection*{DATA AVAILABILITY}
The data that supports the findings of this study are available within the article.

\begin{acknowledgements}
We are grateful to T. Mori and H. Fuketa for the fruitful discussions.
We are also grateful to Y. Yamamoto for his technical support in using SmartSpice. 
This work was partly supported by MEXT Quantum Leap Flagship Program (MEXT Q-LEAP) Grant Number JPMXS0118069228, Japan.
\end{acknowledgements}

\clearpage
\appendix

\section{SET characteristics determined by the orthodox theory}
\begin{figure}
\centering
\includegraphics[width=8.8cm]{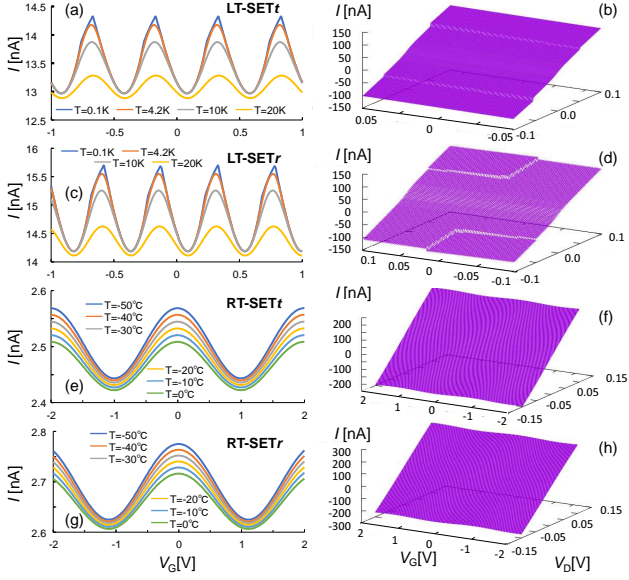}
\caption{
Current-voltage characteristics of SETs calculated based on the orthodox theory. 
(a)(c)(e)(g) Temperature dependence of the $I_{\rm SET}-V_{\rm G}$ characteristics at $V_{\rm D}=0.01$ V. (b)(d) Three-dimensional (3D) view of $I_{\rm SET}$ for $T$ = 0.1 K. 
(f)(g) 3D view of $I_{\rm SET}$ for $T = -30^\circ$C. 
(a)(b)$C_{\rm up}=1$ aF, $C_{\rm dn}=10$ aF, $C_g=2$ aF, $R_{\rm up}=100$ k$\Omega$, $R_{\rm dn}=1$ M$\Omega$ (LT-SET$t$ in Table I).
(c)(d)$C_{\rm up}=1.1$ aF, $C_{\rm dn}=9$ aF, $C_g=2.2$ aF, $R_{\rm up}=110$ k$\Omega$, $R_{\rm dn}=0.9$ M$\Omega$.(LT-SET$r$ in Table I).
As the asymmetry of the two junctions becomes larger, the Coulomb staircase becomes clearer.
(e)(f)$C_{\rm up}=0.1$ aF, $C_{\rm dn}=0.5$ aF, $C_g=0.5$ aF, $R_{\rm up}=100$ k$\Omega$, $R_{\rm dn}=500$ k$\Omega$ (RT-SET$t$ in Table I).
(g)(h)$C_{\rm up}=0.11$ aF, $C_{\rm dn}=0.45$ aF, $C_g=0.45$ aF, $R_{\rm up}=90$ k$\Omega$, $R_{\rm dn}=450$ k$\Omega$ (RT-SET$r$ in Table I).
}
\label{figSETcur}
\end{figure}

Figure~\ref{figSETcur} shows the current-gate voltage characteristics of the four SETs used in the main text, 
calculated based on the orthodox theory~\cite{Amman}. 
The Coulomb staircase becomes clearer for a large asymmetry between the two tunnel junctions; 
however, the amplitude of the oscillation becomes smaller. Therefore, we consider the case of a smaller asymmetry.
The wide-range view of the current–voltage ($I_{\rm D}-V_{\rm D}$) characteristics o
f the SET shown in Figs.~\ref{figSETcur}(b)(d)(f)(h) enables the approximation of 
$V_{\rm D} \propto I_{\rm D}$ of Eq.(2) in the main text.

\begin{figure}
\centering
\includegraphics[width=8.8cm]{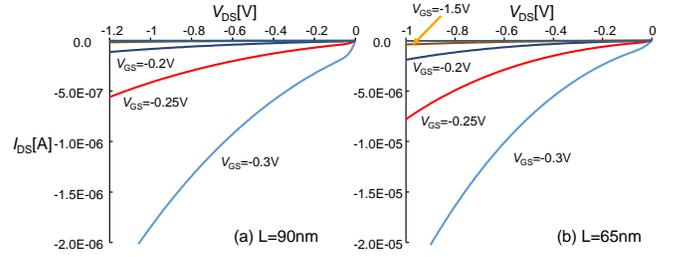}
\caption{
$I_{\rm D}-V_{\rm D}$ characteristics of the pMOS transistors at -30$^\circ$ C.
(a) $L=$90 nm and (b) $L$=65 nm.
}
\label{figpmosbare}
\end{figure}

\section{Additional results for the transistors and amplification circuits}
\begin{figure}
\centering
\includegraphics[width=6cm]{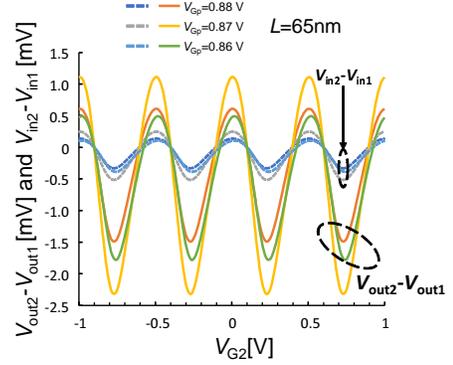}
\caption{
Numerical results of the two outputs $V_{\rm out1}$ and $V_{\rm out2}$ of the 
differential amplifier circuit 
for $L=$65 nm for a fixed $V_{\rm G2}=0$. 
The transistor sizes are the same as those in Fig.~3 in the main text.
}
\label{figdiff65nm}
\end{figure}

Figure~\ref{figpmosbare} shows the basic $I_{\rm D}$-$V_{\rm D}$ characteristics of the pMOS transistors used 
herein (Berkeley short-channel transistor model BSIM-4), 
and Fig.~\ref{figdiff65nm} shows the characteristics of the DA circuits of Fig.~3 (b) for $L$ = 65 nm. 
Although amplification is not observed in the triode region, herein, we show that amplification can be expected
even in the triode region. 
In the triode region, Eq.~(1) is replaced by
\begin{eqnarray}
I_{\rm D} \!&=&\! \beta_p (V_{\rm Gp}-V_{\rm D}-V_{\rm thp}-[V_{\rm D}-V_{\rm out}])[V_{\rm D}-V_{\rm out}], \ \ \ \ \label{eq1-3}
\end{eqnarray}
where $\beta_p\equiv \mu_p C_{\rm ox} \frac{W}{L} $ and $\lambda(<1)$ are the channel length modulation coefficients
($L$, $W$, $C_{\rm ox}$, and $\mu_p$ are the length, width, gate capacitance and mobility of the pMOS, respectively).
The solution to the equations is given by
\begin{equation}
V_{\rm out}\approx V_{\rm D}+2v_{\rm gp} -\frac{1}{R_{\rm D}\beta_p}, \label{eq1-4}
\end{equation}
where $v_{\rm gp}=V_{\rm Gp}-V_{\rm D}-V_{\rm thp}$.
Eq.~(\ref{eq1-4}) shows that the SET output changes the rate of $R_{\rm D} \partial V_{\rm out} / \partial R_{\rm D} \approx 1/[\beta_p R_{\rm D}]$.
When we take 
$\mu_p=0.025$~m${}^2$/(Vs),
$C_{\rm ox}=\epsilon_0 \epsilon/t_{\rm ox}$
with 
$\epsilon=3.9$ and
$t_{\rm ox}=4$~nm
for $L=$90~nm and $W_p=0.5~\mu$m,
we have $\beta_p=0.0002397$ AV$^{-2}$ and $\beta_p R_{\rm D}=2.397\times10^2$~V${}^{-1}$
for $R_{\rm D}=1~$M$\Omega$.
Then the change rate  $R_{\rm D} \partial V_{\rm out}/\partial R_{\rm D} \approx 4.17$~mV. 
Whether the triode region or the saturation region is better for the amplification 
depends on the current characteristics of the MOS transistors.

\section{Temperature dependence}
\begin{figure}
\centering
\includegraphics[width=5.5cm]{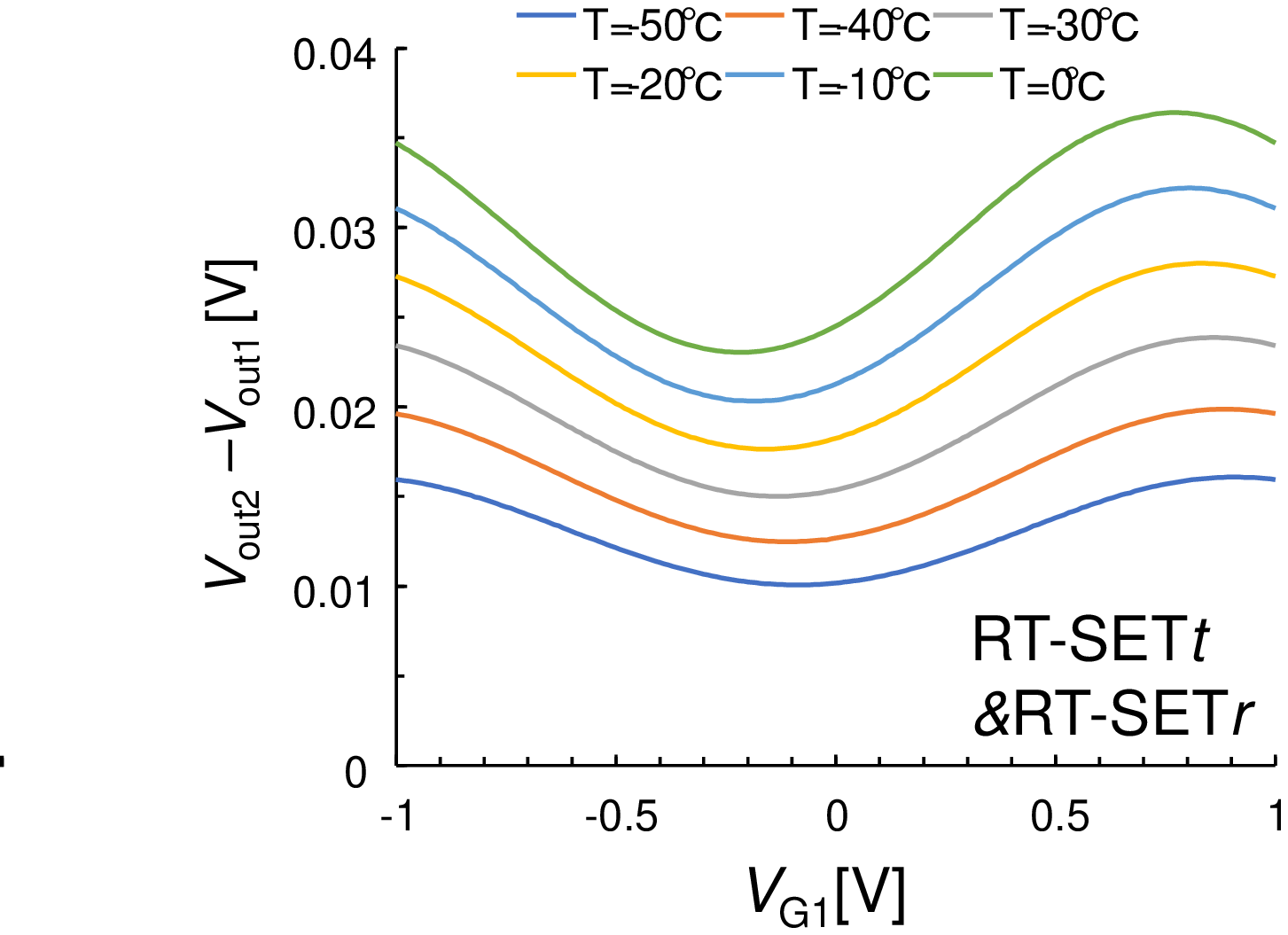}
\caption{
Temperature dependence of $V_{\rm out1}-V_{\rm out2}$
in Fig.~3(a) of $L$=90~nm MOS transistors for different SETs
as a function of $V_{\rm G1}$.  $V_{\rm G2}$=0.1~V.
The left SET is the RT-SET$t$ and the right SET is the RT-SET$r$.
}
\label{figdifftemp}
\end{figure}

\begin{figure}
\centering
\includegraphics[width=5.5cm]{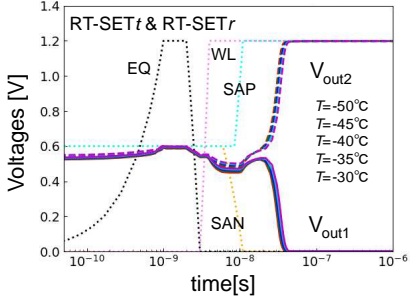}
\caption{
Temperature-dependent characteristics for $V_{\rm G2}=0.2~$V($L$=90~nm). 
The widths of the nMOS and pMOS in the circuits are 0.5 $\mu$m and 0.6 $\mu$m. 
The width of the equalizer nMOS is 20 $\mu$m.
$V_{\rm Gp1}=V_{\rm Gp2}=0.87$~V. 
}
\label{figDRAMtemp}
\end{figure}

We calculated the temperature dependence of the amplifier response. 
It is desirable that the relative magnitude of $V_{\rm out1}$ to $V_{\rm out2}$ does not change 
as long as the operating temperature changes in the range of approximately -50$^\circ \sim 0^\circ$C.
Figures~\ref{figdifftemp} and \ref{figDRAMtemp} show the temperature dependence of the characteristics of the DA (Fig.~3), and DRAM (Fig.~5). 
For example, depending on the value of $V_{\rm Gp}$, 
we can observe the relative change in the magnitude between $V_{\rm out1}$ and $V_{\rm out2}$. 
We also observe that even when there are variations, 
the two SETs can be differentiated by changing the gate voltage $V_{\rm G1}$ or $V_{\rm G2}$. 
The strength to the change of temperature depends on each operation region.

\begin{figure}
\centering
\includegraphics[width=5.6cm]{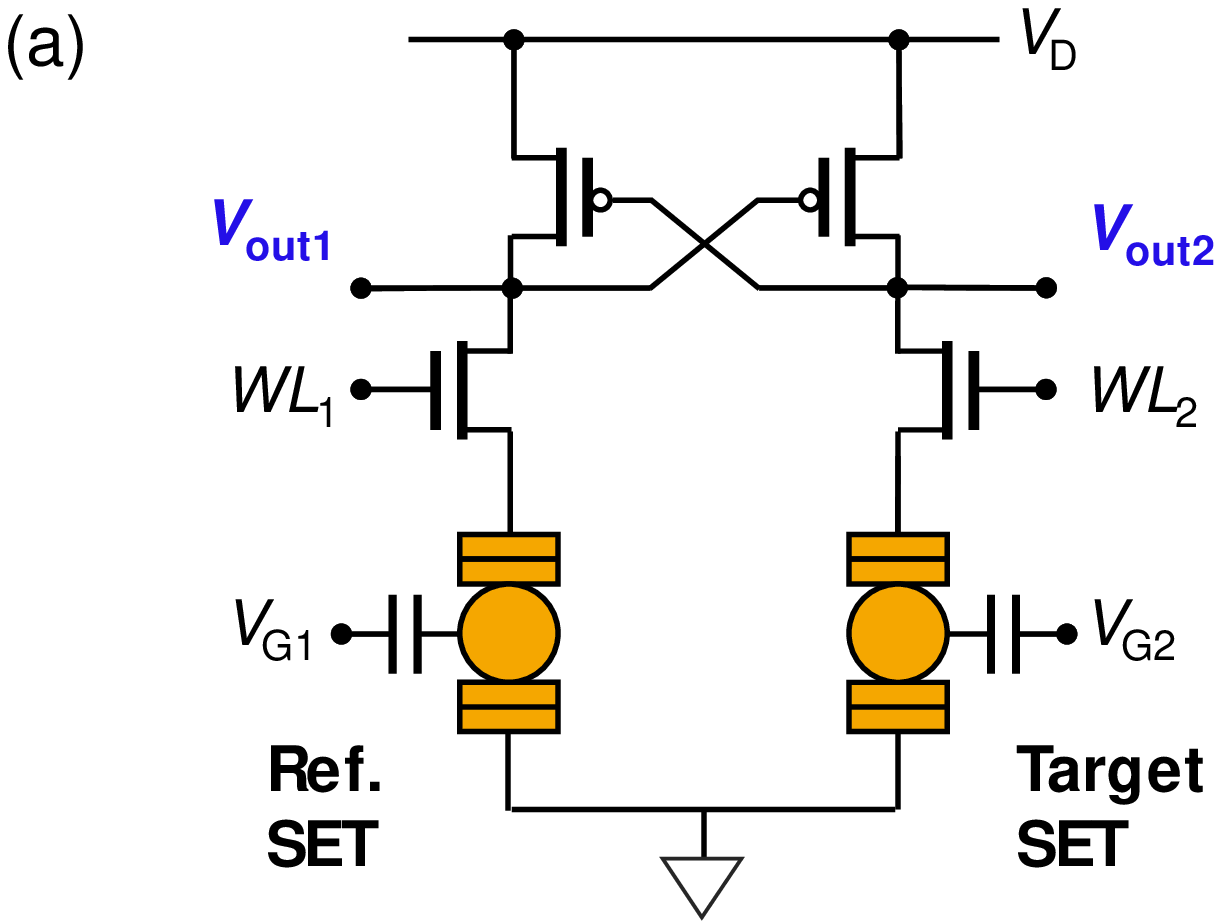}
\includegraphics[width=8.8cm]{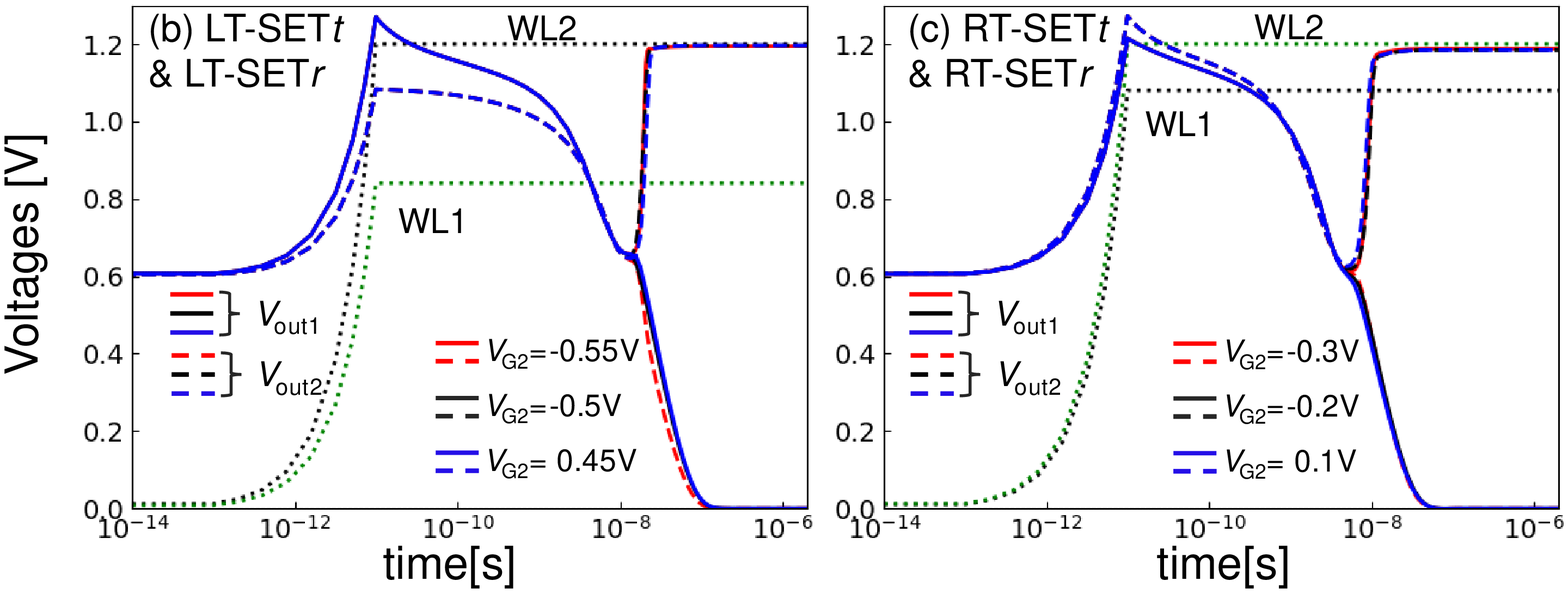}
\caption{
(a) Cross-coupled type readout of two pairs of SETs.
(b)(c) Time-dependent output characteristics of the cross-coupled type
assuming 10 \% variations in the threshold voltage of the MOS transistors 
(the LT-SET pair in (b) and RT-SET pair in (c)).
The two wordline voltages are adjusted to change the relation 
between $V_{\rm G1}$ and  $V_{\rm G2}$ when $V_{\rm G2}$ changes. 
The wordline voltages are switched on at 10 ps with a rising time of 10 ps.
$W_n=5~\mu$m, $W_p=0.1~\mu$m and $V_{\rm G1}$=0~V.}
\label{cross}
\end{figure}

\section{Detection by cross-coupled circuit}
Figure~\ref{cross}(a) shows our other proposal based on the cross-coupled type.
In this case, two SETs are series-connected to cross-coupled MOS transistors,
where the changes in the SET signals are directly transferred to the MOS transistors. 
Herein, the 10\% variations between the SETs and 
among the MOS transistors are considered.
To observe the change in $V_{\rm G2}$ to the fixed $V_{\rm G1}$, 
we have to adjust the wordline voltages. 
In Figs.~\ref{cross}(b) and \ref{cross}(c), the wordline voltage of the left SET is reduced to 
$WL_1\approx 0.84$~V, and $WL_1\approx 1.08$~V, respectively. 
In Fig.~\ref{cross}(b)((c)), $V_{\rm out2}$ becomes higher than $V_{\rm out1}$ between 
$V_{\rm G2}=-0.55$~V and  $V_{\rm G2}=-0.5$~V ($V_{\rm G2}=-0.3$~V and  $V_{\rm G2}=-0.2$~V).
When there are no variations, the left and right hand circuits are
exactly the same, and $V_{\rm out1}$ and $V_{\rm out2}$ are clearly split 
depending on the gate voltages of the two SETs.
Temperature changes are not observed, similar to the case with the SRAMs mentioned above (Figures not shown).


\end{document}